\documentclass[aps,prl,reprint,superscriptaddress]{revtex4-1}

\usepackage{graphicx}
\usepackage{amssymb, amsmath}
\usepackage[usenames]{color}

\begin{document}

\preprint{}

\title{Tensile Strained Gray Tin: a New Dirac Semimetal for Observing Negative Magnetoresistance with Shubnikov-de-Haas Oscillation}

\author{Huaqing Huang}
\affiliation{Department of Materials Science and Engineering, University of Utah, Salt Lake City, Utah 84112, USA}

\author{Feng Liu\footnote{Corresponding author: fliu@eng.utah.edu}}
\affiliation{Department of Materials Science and Engineering, University of Utah, Salt Lake City, Utah 84112, USA}
\affiliation{Collaborative Innovation Center of Quantum Matter, Beijing 100084, China}

\date{\today}

\begin{abstract}
The extremely stringent requirement on material quality has hindered the investigation and potential applications of exotic chiral magnetic effect in Dirac semimetals. Here, we propose that gray tin is a perfect candidate for observing the chiral anomaly effect and Shubnikov-de-Haas (SdH) oscillation at relatively low magnetic field. Based on effective $k.p$ analysis and first-principles calculations, we discover that gray tin becomes a Dirac semimetal under tensile uniaxial strain, in contrast to a topological insulator under compressive uniaxial strain as known before. In this newly found Dirac semimetal state, two Dirac points which are tunable by tensile [001] strains, lie in the $k_z$ axis and Fermi arcs appear in the (100) surface. Duo the low carrier concentration and high mobility of gray tin, a large chiral anomaly induced negative magnetoresistance and a strong SdH oscillation are anticipated in this half of strain spectrum. Comparing to other Dirac semimetals, the proposed Dirac semimetal state in the nontoxic elemental gray tin can be more easily manipulated and accurately controlled. We envision that gray tin provides a perfect platform for strain engineering of chiral magnetic effects by sweeping through the strain spectrum from positive to negative and vice versa.
\end{abstract}


\maketitle

The discovery of Dirac and Weyl semimetals with chiral quasiparticles \cite{Na3Bi,Cd3As2,TaAsHLin,TaAsPRX,huanghqPtSe2} opens a new avenue to realizing the long-anticipated high-energy-physics Adler-Bell-Jackiw chiral anomaly\cite{adler1969axial,bell1969pcac,nielsen1983adler} in condensed matter systems. As a defining signature, a large negative longitudinal magnetoresistance (MR) is expected to be observable in Dirac semimetals. In the presence of parallel magnetic and electric fields, each Dirac point split into two Weyl nodes with opposite chirality. The Weyl fermions residing at one Weyl node are pumped to the other, resulting in non-conserved chiral charges and prominent negative MR. The chiral anomaly induced negative MR need to be observed in Dirac semimetals with ultralow carrier concentration and high mobility, which require a high sample quality. Apart from the negative MR effect, the Shubnikov-de-Haas (SdH) oscillation is another interesting magnetotransport phenomenon in Dirac semimetals, where the MR oscillates periodically in reciprocal magnetic field ($1/B$).
Analysis of the SdH oscillations of MR gives a nontrivial $\pi$ Berry phase, which is a distinguished feature of Dirac fermions.\cite{PhysRevLett.113.246402,cao2015landau} This quantum oscillation is attributed to the Dirac band structure and ultrahigh carrier mobility of Dirac materials.

However, to date, the chiral anomaly induced negative MR was observed only in a few Dirac semimetal materials\cite{li2015giant,*li2016negative,xiong2015evidence,li2016chiral}, and none of which exhibits negative MR associated with a SdH oscillation in the quantum limit. This is due to the lack of a high-quality Dirac semimetal with low carrier concentration and high carrier mobility at the same time.
For example, the negative MR without SdH oscillation was observed in Cd$_3$As$_2$ and Na$_3$Bi with low carrier concentrations ($\sim10^{17}$ cm$^{-3}$) and mobilities ($\sim10^3$ cm$^2$V$^{-1}$s$^{-1}$)\cite{li2015giant,*li2016negative}. Although an ultrahigh carrier mobility of $\sim 10^6$ cm$^2$V$^{-1}$s$^{-1}$ that accompanied by strong SdH oscillations was reported in different Cd$_3$As$_2$ samples, the chiral anomaly induced negative MR was not observed due to the relatively high carrier concentration ($10^{18}- 10^{19}$cm$^{-3}$)\cite{liang2015ultrahigh}. In addition, Na$_3$Bi is unstable and decomposes rapidly upon exposure to air. The delicate synthesis together with the high toxicity of Cd and As also makes handling of Cd$_3$As$_2$ difficult. Given the above challenges, identifying new Dirac semimetals with low carrier concentration and high carrier mobility is of great importance for both basic research and potential applications.

In this Letter, we predict that tensile strained gray tin is a Dirac semimetal, and more importantly propose it to be a perfect candidate to observe the chiral anomaly effect and SdH oscillation at relatively low magnetic field. So far it has been taken for granted that external strains just drive gray tin into a topological insulator state \cite{LiangFu,strainSnTI,*strainSnTI2}. Here, we reveal the missing half of the strain spectrum of gray tin, where a previously unknown Dirac semimetal phase is discovered. Based on effective $k\cdot p$ analysis and first-principles calculations, we demonstrated the existence of a pair of controllable Dirac points in the $k_z$ axial of the bulk Brillouin zone (BZ) and Fermi arcs connecting the two projected Dirac points on the (100) surface when gray tin is under tensile uniaxial strain.
Due to the relatively low carrier concentration \cite{PhysRev.134.A1261,PhysRev.122.1431} and anomalously high mobility \cite{ewald1959, broerman1970anomalous,*broerman1971effect}, a large chiral anomaly induced negative MR accompanied by SdH oscillation is expected to be observable in the tensile strained gray tin. In fact, some measurements of gray tin many years ago have addressed the negative MR effect associate with the SdH oscillation manifesting an unconventional Berry phase \cite{PhysRev.134.A1261,PhysRev.122.1431}, which supports our proposal. Comparing to other Dirac semimetals, the proposed Dirac semimetal state in the nontoxic elemental gray tin promises ease of fabrication and tuning. Our findings provide a perfect platform for strain engineering of chiral magnetic effect by scanning through the strain spectrum from positive to negative and vice versa.

Gray tin, also known as the $\alpha$ phase of Sn crystal ($\alpha$-Sn), is a common zero-gap semiconductor, in which the conduction band minimum and valance band maximum are degenerate at the $\Gamma$ point \cite{PhysRevLett.11.194,PhysRevB.2.352}.
Gray tin crystallizes in the diamond structure with $Fd\bar{3}m$ symmetry (space group No. 227) as shown in Fig. 1(a).
Unlike other group-IV compounds with the same crystal structure, $\alpha$-Sn has an inverted band ordering where the \textit{p}-orbital derived $\Gamma^+_8$ state located at the Fermi level is higher than the \textit{s}-orbital dominated $\Gamma^-_7$ state with opposite parity [see middle panel of Fig.1(c)].
The band inversion is critical for the existence of topologically nontrivial states including topological insulator and Dirac semimetal phases. Since the fourfold degeneracy of the $\Gamma_8^+$ state is a consequence of the $O_h$ symmetry of the diamond lattice, applying an uniaxial strain will split the degenerate states. However, we discover that the split manifests in different ways depending on the direction of the applied strain, which lead to different topological states.

\begin{figure}
\includegraphics[width =1\columnwidth]{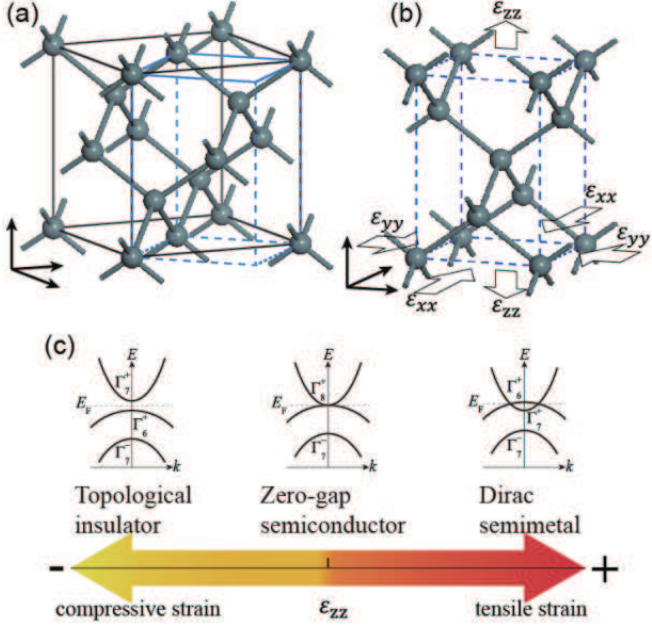}%
\caption{\label{fig1_stru} (Color online) (a) Crystal structure of $\alpha$-Sn with $Fd\bar{3}m$ (No. 227) symmetry in a cubic unit cell. The dashed lines indicate a tetragonal unit cell. (b) Tetragonal unit cell under external strains. Substrate induced strains are applied as in-plane compressive strains leading to a uniaxial $z$-axis tensile strain. (c) Topological phase diagram v.s. uniaxial $z$-axis strain. For tensile strain, $\varepsilon_{zz}>0$, the system becomes a Dirac semimetal with two Dirac points; while for compressive strain, $\varepsilon_{zz}<0$, the system is a topological insulator.}
\end{figure}

We first perform a $k\cdot p$ analysis to investigate general behavior of electronic structures under strain. The $\Gamma_8^+$ bands at the $\Gamma$ point are fourfold degenerate as $J=3/2$ multiplet, which can be effectively described by the $k\cdot p$ Hamiltonian\cite{luttinger,kane,bir1974symmetry}: \begin{equation}
H(\mathbf{k})=-(\gamma_1+\frac{5}{2}\gamma_2)k^2+2\gamma_2 \sum_{i}k_i^2J_i^2+2\gamma_3\sum_{i\neq j}k_ik_j\{J_iJ_j\},
\end{equation}
where $i,j=(x,y,z)$, $J_i$ are $4\times4$ spin-3/2 matrices, $\{J_iJ_j\} = \frac{1}{2}(J_iJ_j + J_jJ_i)$. The inverse-mass parameters are $\gamma_{1}=-19.2, \gamma_2=-13.2$ and $\gamma_3=-8.9$ (in unit of $\hbar^2/2m_e$), which have been determined by Booth \textit{et al.} from an experimental study of the anisotropy of the MR oscillations in gray tin \cite{PhysRev.168.805}.
The main perturbations induced by strains are given by
\begin{equation}
H_{\varepsilon}=(a+\frac{5}{4}b)\varepsilon-b\sum_iJ_i^2\varepsilon_{ii}-\frac{1}{\sqrt{3}}d\sum_{i\neq j}\{J_iJ_j\}\varepsilon_{ij},
\end{equation}
where $\varepsilon=\varepsilon_{xx}+\varepsilon_{yy}+\varepsilon_{zz}$. The deformation potentials of $b=-2.3$ eV and $d=-4.1$ eV are defined analogously to the inverse-mass parameters, which are determined experimentally too \cite{price1969crystal,PhysRevB.5.3914}. The eigenvalues of the effective $k\cdot p$ Hamiltonian $H_{k\cdot p}=H(\mathbf{k})+H_{\varepsilon}$ for strained $\alpha$-Sn can be derived analytically, which are given in the Supplemental Material \footnote{\label{fn}See Supplemental Materials at http://link.aps.org/supplemental/xxx, for more details about the $k\cdot p$ analysis, the computational methods and the estimation of longitudinal MR of tensile strained $\alpha$-Sn, which include Ref.~\cite{PBE,heyd2003,*heyd2004,lopez,*lopez2,wannier90}}.

For simplicity, let's first consider an uniaxial [001] strain, i.e., $\varepsilon_{xx}=\varepsilon_{yy}\neq\varepsilon_{zz}$ and $\varepsilon_{xy}=\varepsilon_{yz}=\varepsilon_{xz}=0$ \cite{PhysRevB.12.2336,averous1979symmetry}. In this case, the cubic $Fd\bar{3}m$ (No. 227, $O_h^7$) symmetry reduces to tetragonal $I4_1/amd$ (No. 141, $D_{4h}^{19}$). The $\Gamma_8^+$ state splits into $\Gamma_6^+$ and $\Gamma_7^+$, and the corresponding eigenvalues along $k_z$ axis become,
\begin{eqnarray}
E_{1,2}(k_z)&=&-\gamma_1k_z^2+a\varepsilon \nonumber\\
&\pm&\left[4\gamma_2^2k_z^4+2\gamma_2b(\varepsilon-3\varepsilon_{zz})k_z^2+\frac{b^2}{4}(\varepsilon-3\varepsilon_{zz})^2\right]^{\frac{1}{2}} \nonumber\\
&=&-\gamma_1k_z^2+a\varepsilon\pm\left|2\gamma_2k_z^2+\frac{b}{2}(\varepsilon-3\varepsilon_{zz})\right|.
\end{eqnarray}
Apparently, the criterion for the existence of band crossing points in $k_z$ axis is given by the following condition:
\begin{equation}
\gamma_2b(\varepsilon-3\varepsilon_{zz})<0.
\label{eq_crit}
\end{equation}

We can apply either a tensile $z$-axis strain or a biaxial compressive in-plane $(x,y)$ strain, $\varepsilon_{zz}>0$, whereas $(\varepsilon-3\varepsilon_{zz})=\varepsilon_{xx}+\varepsilon_{yy}-2\varepsilon_{zz}<0$. Inserting the parameters given above, we found that $\gamma_2b>0$, and the criterion of Eq.~(\ref{eq_crit}) is satisfied. As a consequence, the system becomes a Dirac semimetal with two symmetric Dirac points at $(0,0,k_z^c=\pm\sqrt{b(3\varepsilon_{zz}-\varepsilon)/4\gamma_2})$, which are obviously tunable by external strain $\varepsilon_{zz}$. This Dirac semimetal phase, which was previously overlooked, reveals the missing half of the strain spectrum on $\alpha$-Sn and also suggests exotic properties of other strained materials with similar electronic structures, such as HgTe \cite{HJZhangHgTe,tensileHgTe111}. In contrast, when applying a compressive [001] strain or a tensile biaxial in-plane strain, $\varepsilon_{zz}<0$, $\gamma_2b(\varepsilon-3\varepsilon_{zz})>0$, the Dirac points vanish and the system is driven to a topological insulator with a finite energy gap, as shown before\cite{PhysRevB.5.3914,LiangFu,strainSnTI,*strainSnTI2}.
Based on the $k\cdot p$ analysis, we obtain the phase diagram of $\alpha$-Sn under a [001] strain, as schematically shown in Fig. 1(c). A topological phase transition from topological insulator to Dirac semimetal can be driven by tuning $\varepsilon_{zz}$ through zero. We also studied the effect of [111] strain, which show similar topological phase transition \footnotemark[\value{footnote}].

\begin{figure}
\includegraphics[width =0.9\columnwidth]{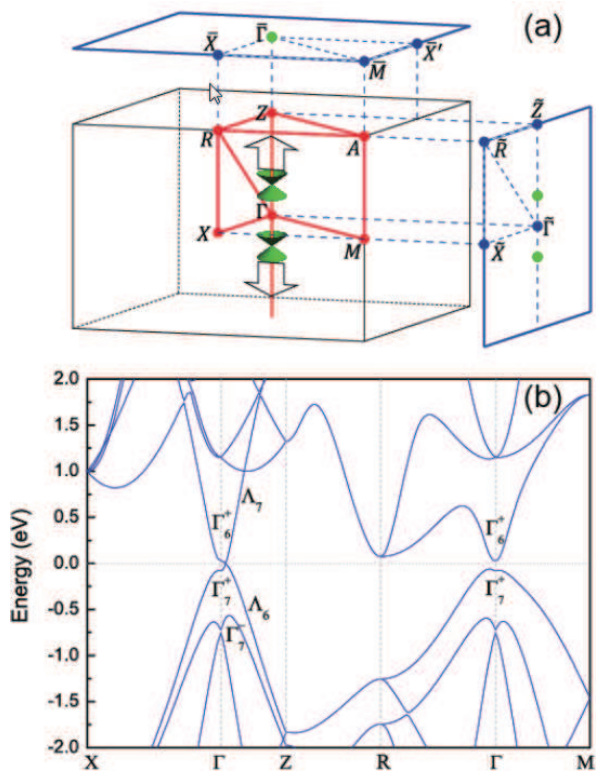}%
\caption{\label{fig2_band} (Color online) (a) The Brillouin zone of bulk $\alpha$-Sn in tetragonal unit cells and the projected surface Brillouin zones of (001) and (010) surfaces. (b) Calculated HSE band structure of $\alpha$-Sn under in-plane compressed strain of $-1\%$.}
\end{figure}

To verify the prediction from the effective $k\cdot p$ analysis presented above, we perform first-principles calculations on strained gray tin using Vienna \textit{ab initio} simulation package\cite{VASP}.
More details of models and computational methods are presented in the Supplemental Material \footnotemark[\value{footnote}].
For a biaxial in-plane compressive strain of $-1\%$, the $\Gamma_8^+$ state splits into $\Gamma_7^+$ and $\Gamma_6^+$, which are pushed down and up, respectively. Since the $\Lambda_7$ and $\Lambda_6$ bands belong to different irreducible representations and disperse upward and downward respectively along the $\Gamma$-A line, the two bands cross at two discrete points: (0, 0, $\pm 0.107$) (in unit of $\pi/a$), consistent with the Dirac points predicted by the effective $k\cdot p$ theory, as shown in Fig.~\ref{fig2_band}. The Fermi level is exactly at the band crossing points which are fourfold degenerate due to the coexistence of time-reversal and inversion symmetries. Thus the Fermi surface consists of two isolated points, around which the bands disperse linearly, resulting in a 3D Dirac semimetal. The separation of the two Dirac points in momentum space increases with the increasing external strain. If the tensilel strain is too large, the conduction band at the $R$ point would shift concave downwards and become occupied. To satisfy the charge neutral condition, the Fermi level would diverge from the Dirac point. In contrast, by applying a compressive [001] strain, the $\Gamma_7^+$ and $\Gamma_6^+$ states are lifted in opposite direction, and a globe band gap opens in the entire BZ (not shown). As the band inversion retains in the compressively strained $\alpha$-Sn, this gapped system is a topological insulator\cite{LiangFu}.

Because the splitting of $\Gamma_8^+$ does not change the band inversion in $\alpha$-Sn, the nontrivial topology of the Dirac semimetal state under tensile strain should be similar to the topological insulator state under compressive strain. Also, the band structures are gapped in both the $k_z=0$ and the $k_z=\pi$ planes when the system is under compressive or tensile strains, the $\mathbb{Z}_2$ topological invariants in these planes are well-defined. In fact, as inversion symmetry retains in the strained system, we can simply determine the $\mathbb{Z}_2$ index from the parities of all occupied bands at time-reversal invariant momentum (TRIM)\textbf{ k}-points\cite{LiangFu}. The parity products of occupied bands is $-1$ at $\Gamma$ and $+1$ at other TRIMs, hence the $\mathbb{Z}_2=1$ for the $k_z=0$ plane, whereas $\mathbb{Z}_2=0$ for the $k_z=\pi$ plane. Therefore, the strained $\alpha$-Sn is always topologically nontrivial. Thus, topological surface states or Fermi arcs are expected to appear on side surfaces of the compressively or tensile strained gray tin.

\begin{figure}
\includegraphics[width =1\columnwidth]{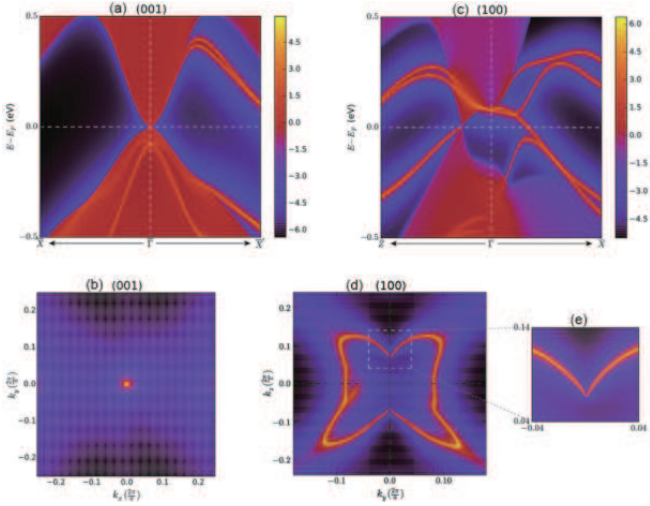}%
\caption{\label{fig3_ss} (Color online) The projected surface states and corresponding Fermi surface of semi-infinite $\alpha$-Sn under a compressive in-plane strain of $-1\%$. (a,b) and (c-e) The projected surface local density of states and Fermi arcs for the (001) and (100) surfaces, respectively.}
\end{figure}

One of the most important consequences of Dirac semimetal is the existence of topological surface states and Fermi arcs on the surface. We have calculated both the (001) and (010) surface states of tensile strained $\alpha$-Sn, as shown in Fig.~\ref{fig3_ss}. For the (001) surface, two Dirac points are projected to the same point of the surface BZ. Bulk continuum superimposes nontrivial surface states, and the Fermi surface of the (001) surface is just a single point [Fig. 3(b)]. For the (010) surface, even though there are some trivial surface bands due to the dangling bond states of the unsaturated surface Sn atoms, the nontrivial surface states, which originate from the gapless point, are clearly visible [see Fig. 3(c)]. As shown in Fig. 3(d), the Fermi surface, which has a shape of butterfly, is composed of two pieces of Fermi arcs, which connect the two projections of bulk Dirac points. However the Fermi velocity is ill-defined at these projected Dirac points [i.e., singular points, see Fig. 3(e)]. Although the Fermi arc pattern may change upon varying surface potential, its existence, which stems from the bulk 3D Dirac points, is robust against such perturbations. These unique features, absent for topological insulators, can be measured by angle-resolved photoemission spectroscopy techniques.

This newly discovered Dirac semimetal phase in strained $\alpha$-Sn is expected to facilitate the realization of the Adler-Bell-Jackiw chiral anomaly\cite{adler1969axial,nielsen1983adler}, which is observable as a negative longitudinal MR. To do so, it is required that the carrier density is low enough so that the Fermi level is located close to the Dirac point. This condition is clearly satisfied by $\alpha$-Sn with a known low carrier concentration on the order of $10^{16}$  cm$^{-3}$\cite{PhysRev.134.A1261,PhysRev.122.1431}, Moreover, as the mobility of $\alpha$-Sn is anomalously high ($\sim 10^5$cm$^2$V$^{-1}$s$^{-1}$, comparable to that of the high-mobility Dirac semimetal Cd$_3$As$_2$) and increases dramatically with decreasing carrier concentration \cite{ewald1959, broerman1970anomalous,*broerman1971effect}, it is easy to drive the system into the extreme quantum limit at relatively low magnetic field. In fact, some measurements many years ago have shown the negative MR effect and the SdH oscillations with an anomalous oscillatory phase of $-\pi/2$, which indicate strong signatures of the Adler-Bell-Jackiw chiral anomaly in gray tin \cite{PhysRev.134.A1261,PhysRev.122.1431}. In addition, a giant non-saturating linear transverse MR is expected in strained $\alpha$-Sn, which can be useful to clarify the unclear mechanism for the linear MR in Dirac materials.

\begin{figure}
\includegraphics[width =1\columnwidth]{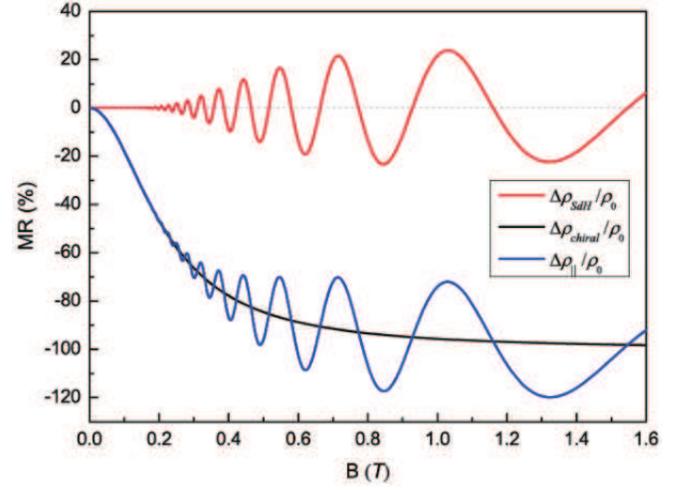}%
\caption{\label{fig4_mr} (Color online) The estimated longitudinal MR as a function of magnetic field at 1.2 K. The SdH oscillation term $\Delta\rho_{SdH}/\rho_0$, the chiral anomaly induced negative MR $\Delta\rho_{chiral}/\rho_0$ and the total MR $\Delta\rho_{\parallel}/\rho_0$ are shown in red, black and blue, respectively}
\end{figure}

To further assess the chiral anomaly induced negative MR and SdH oscillation with nontrivial Berry phase, the behavior of longitudinal MR in strained gray tin is simulated. When an external electric field \textbf{E} is applied in parallel with the magnetic field \textbf{B}, the chiral charges at one node are pumped to the other with opposite chirality due to the chiral anomaly induced $\pm\frac{e^3}{4\pi^2\hbar^2}\mathbf{E}\cdot\mathbf{B}$ term. This charge pumping yields a positive magnetic conductivity (correspond to a negative MR) given by\cite{PhysRevLett.113.247203},
\begin{equation}
\Delta \sigma_{chiral}=\frac{e^4\tau_a}{4\pi^4\hbar^4g(E_F)}B^2
\end{equation}
where $g(E_F)$ is the density of state (DOS) at the Fermi energy $E_F$, $\tau_a$ is the internode scattering time.
Meanwhile, due to the high mobility of strained gray tin, the quantum oscillation of the MR are expected to be observed at low temperature, which can be described by the Lifshitz-Kosevich formula\cite{magneticoscillation}:
\begin{equation}
\frac{\Delta \rho_{SdH}}{\rho_0}=A(T,B)\cos\left[2\pi(\frac{F}{B}-\gamma\pm\frac{1}{8})\right].
\end{equation}
The oscillatory phase factor $2\pi\gamma=\pi-\varphi_B$ is directly related to the Berry phase $\varphi_B$. A nontrivial $\pm\pi$ Berry phase can be acquired by electrons in cyclotron orbits.
We estimated the longitudinal MR curve of a strained gray tin with the carrier concentration of $n=2.0\times10^{16}$ cm$^{-3}$ and the mobility of $\mu=2.5\times 10^5$ cm$^2$V$^{-1}$s$^{-1}$, which are in the experimentally accessible range\cite{PhysRev.134.A1261,PhysRev.122.1431}. As shown in Fig.~\ref{fig4_mr}, the oscillatory MR $\Delta\rho_{\parallel}/\rho$ decreases rapidly with the magnetic field as expected. The chiral anomaly induced negative MR $\Delta\rho_{chiral}/\rho_0$ can approach to $-100\%$ with a weak magnetic field. This implies the major contribution to the total MR ${\Delta\rho_{\parallel}}/{\rho_0}$ from the chiral anomaly. Due to the small cross-sectional area $A_F$ of the Fermi surface, the estimated oscillation frequency $F$ is only about $2.3$ T according to the Onsager relation $F=A_F\hbar/2e\pi$,
much smaller than other Dirac semimetals. These novel behaviors of MR, is rare in non-ferromagnetic materials, thus can serve as one of the most definite signatures of the Dirac semimetal state in strained $\alpha$-Sn (More details about the estimation are presented in Supplemental Materials\footnotemark[\value{footnote}]).

In conclusion, we discover a Dirac semimetal state in the other missing half of the tensile strain spectrum of gray tin, which offers a perfect candidate for the realization of chiral magnetic effects, addressing a long-standing experimental challenge. The exotic chiral anomaly induced large negative longitudinal MR associated with SdH oscillation is estimated.
Furthermore, gray tin also provides a new route to studying the interplay between different topological states and other novel phenomena.
For example, Weyl semimetals are hopefully realized in gray tin by breaking either time-reversal or inversion symmetries. Specifically, Weyl points are expected to be obtained by splitting Dirac points in strained $\alpha$-Sn via doping, alloying, and straining, which will be discussed in future work.

\begin{acknowledgments}
This work was support by DOE-BES (Grant No. DE-FG02-04ER46148)
\end{acknowledgments}

\providecommand{\noopsort}[1]{}\providecommand{\singleletter}[1]{#1}%

\end{document}